\documentclass[lettersize,journal]{IEEEtran}
\usepackage{amsmath,amssymb,amsfonts}
\usepackage{algorithmic}
\usepackage{algorithm}
\usepackage{array}
\usepackage[caption=false,font=normalsize,labelfont=sf,textfont=sf]{subfig}
\usepackage{textcomp}
\usepackage{stfloats}
\usepackage{url}
\usepackage{verbatim}
\usepackage{graphicx,color}
\usepackage{cite}
\usepackage{xcolor}
\begin{document}

\title{SynthSoM-Twin: A Multi-Modal Sensing-Communication Digital-Twin Dataset for Sim2Real Transfer via Synesthesia of Machines}

\bstctlcite{BSTcontrol}

\author{Junlong Chen,~\IEEEmembership{Graduate Student Member,~IEEE}, Ziwei Huang,~\IEEEmembership{Member,~IEEE}, Xuesong Cai,~\IEEEmembership{Senior Member,~IEEE}, Xiang Cheng,~\IEEEmembership{Fellow,~IEEE},  Liuqing Yang,~\IEEEmembership{Fellow,~IEEE}
}



\maketitle

\begin{abstract}
This paper constructs a novel multi-modal sensing-communication digital-twin dataset, named SynthSoM-Twin, which is spatio-temporally consistent with the real world, for Sim2Real transfer via Synesthesia of Machines (SoM). To construct the SynthSoM-Twin dataset, we propose a new framework that can extend the quantity and missing modality of existing real-world multi-modal sensing-communication dataset. Specifically, we exploit multi-modal sensing-assisted object detection and tracking algorithms to ensure spatio-temporal consistency of static objects and dynamic objects across real world and simulation environments. The constructed scenario is imported into three high-fidelity simulators, i.e., AirSim for generating non-radio frequency (non-RF) sensing data, WaveFarer for generating radio frequency (RF) sensing data, and Sionna RT for generating RF channel data. The SynthSoM-Twin dataset contains spatio-temporally consistent data with the real world, including 66,868 snapshots of synthetic RGB images, depth maps, light detection and ranging (LiDAR) point clouds, millimeter wave (mmWave) radar point clouds, and large-scale and small-scale channel fading data. To validate the utility of SynthSoM-Twin dataset, we conduct Sim2Real transfer investigation by implementing two cross-modal downstream tasks via cross-modal generative models (CMGMs), i.e., cross-modal channel generation model and multi-modal sensing-assisted beam generation model. Based on the downstream tasks, we explore the threshold of real-world data injection that can achieve a decent trade-off between real-world data usage and models' practical performance. Experimental results show that the model training on the SynthSoM-Twin dataset achieves a decent practical performance, and the injection of real-world data further facilitates Sim2Real transferability. Based on the SynthSoM-Twin dataset, injecting less than 15\% of real-world data can achieve similar and even better performance compared to that trained with all the real-world data only. This indicates that a proper digital-twin dataset can save over 85\% of the measurement cost on average.
\end{abstract}



\section{INTRODUCTION}\label{sec:introduction}
\IEEEPARstart{T}{he} integrated sensing and communication (ISAC) is a typical application scenario for sixth-generation (6G) and therefore has been widely investigated~\cite{wen2024, Luong2025}. However, ISAC focuses on the integration of radio frequency (RF) communications and RF sensing~\cite{Luong2025,du2025toward}. In this case, ISAC has difficulty in obtaining the benefits brought by multi-modal information. To adequately utilize multi-modal sensing-communication data, Synesthesia of Machines (SoM)~\cite{cheng2023intelligent}, i.e., artificial intelligence (AI)-native intelligent multi-modal sensing-communication integration, was proposed as a novel concept and framework. Similar to how humans sense the environment via multiple organs, in SoM, multi-modal sensors, including RF and non-RF sensors, together with RF devices, can obtain environmental information to achieve mutual enhancement between communications and multi-modal sensing~\cite{huang2024lidaraid}. As the performance bound of AI-native intelligent multi-modal sensing-communication integration, are determined by the quantity and quality of the dataset, constructing massive and high-quality multi-modal sensing-communication datasets is of paramount importance~\cite{salem2025data,huang2025llm4mg,Gizzini2025ttt}. 




In general, existing multi-modal sensing-communication datasets can be divided into synthetic datasets, real-world datasets, and digital-twin datasets. 

\begin{enumerate}
    \item Synthetic datasets: Synthetic datasets can be generated at low cost and high diversity with massive multi-modal sensing and communication data by leveraging high-fidelity simulators~\cite{Huang2025SoM-ScatRec}. ViWi~\cite{alrabeiah2020viwi} offered synthetic visual sensing data and channel data, conducting research on vision-aided beam generation and blockage prediction. More recently, we constructed a synthetic multi-modal sensing-communication dataset in~\cite{cheng2025synthsom}, named SynthSoM. The SynthSoM dataset contained synthetic RGB images, depth maps, light detection and ranging (LiDAR) point clouds, millimeter wave (mmWave) radar point clouds, and channel data. As the synthetic data is generated from simulators, its fidelity is limited, thus constraining downstream models' practical performance when deployed in real world.
    \item Real-world datasets: Real-world datasets can provide high-quality data captured from real world~\cite{cai2024toward}. RFSAC~\cite{ji2023networking} included real-world RF sensing data and RF channel data, conducting research on RF sensing-assisted communications. To further include multi-modal sensing data, DeepSense 6G~\cite{alkhateeb2023deepsense} included real-world multi-modal sensing and communication data, which can support research on multi-modal sensing-assisted communications. To further focus on high-mobility scenarios, the authors in~\cite{alkhateeb2025deepsensev2v} constructed a real-world vehicle-to-vehicle (V2V) multi-modal sensing-communication dataset, named DeepSense-V2V, which can further support research on multi-modal sensing-assisted communications in V2V scenarios. Compared to synthetic datasets, real-world datasets have higher collection cost and narrower coverage across geographies, traffic patterns, and hardware configurations. As a result, real-world dataset cannot satisfy the large-scale and high diversity of data demands~\cite{huang2018big}. 
    \item Digital-twin datasets: To address the limitations of synthetic datasets and real-world datasets, a straightforward approach is to achieve complementary advantages by properly integrating these two types of datasets. To leverage the advantages of both synthetic datasets and real-world datasets, it can be achieved by constructing a digital-twin dataset based on the digital-twin technology. Digital-twin datasets refer to the simultaneous collection of real-world data from the real world and the creation of a digital replica of the real world where the data is generated, with synthetic data generated within the constructed simulation environments~\cite{wang2022mobility,khan2022digital}. By calibrating the spatial structure, textures, equipment parameters, electromagnetic environment, and other settings of the digital-twin scenario via real-world data, the digital-twin scenario can maintain consistency with the corresponding physical scenario. Therefore, the distribution differences between the data generated from the digital-twin scenario and the real-world data are similar, which enables fewer real-world data to achieve decent models' practical performance~\cite{shoukat2024autonomous,bariah2024interplay}. Some preliminary works have utilized the digital-twin technology to construct digital-twin datasets. SMART~\cite{muruganandham2025smart} included synthetic sensing-communication data that are aligned with the real-world scenario. However, the sensing data solely contained RGB images and LiDAR point clouds, and the communication data was limited to relative beam power. To further include RF sensing data, DeepVerse 6G~\cite{demirhan2025deepverse} offered synthetic RF sensing data, non-RF sensing data, and communication data aligned with the real-world scenario, whereas communication data also contained relative beam power without channel multipath data. Although the SMART dataset~\cite{muruganandham2025smart} and DeepVerse 6G dataset~\cite{ demirhan2025deepverse} consider static scenarios that are consistent with the real world and include synthetic sensing-communication data that are aligned with the real-world scenario, they ignore dynamic objects, e.g., vehicles and pedestrians, which affect both sensing data and RF channels. As a result, the SMART dataset~\cite{muruganandham2025smart} and DeepVerse 6G dataset~\cite{ demirhan2025deepverse} fail to adequately achieve spatio-temporal consistency across real world and simulation environments. 
\end{enumerate}
 As a result, a new multi-modal sensing-communication digital-twin dataset, which achieves spatio-temporal consistency of both static objects and dynamic objects across real world and simulation environments, is urgently required.


Another critical aspect of multi-modal sensing-communication digital-twin dataset is to validate its utility in reducing real-world data usage. Fortunately, we can validate the utility of digital-twin dataset by conducting Sim2Real transfer. Sim2Real transfer refers to training models in simulation and migrating them to the real world. A typical Sim2Real approach is pre-training a model on massive synthetic data and then fine-tuning it with a small amount of real-world data. By conducting Sim2Real transfer, AI models can reduce the performance gap from simulation to real world and enhance Sim2Real transferability, thus reducing real-world data usage~\cite{jeong2020self}. In the 2010s, researchers investigated Sim2Real transfer in the area of robotics and computer vision. For example, the authors in \cite{tobin2017domain} employed domain randomization to generate synthetic data, increasing the diversity of simulation data to enhance Sim2Real transferability for robotic grasping. To further consider data alignment, Park et al. aligned synthetic data with real-world data and applied it to Sim2Real visual grasping, which can improve downstream generalization \cite{park2021sim}. To quantify the sources and contributing factors of the Sim2Real gap, Huch et al. provided a measurable methodology for exploring Sim2Real transferability in LiDAR modality \cite{huch2023quantifying}. To further investigate in AI model training process, Shumailov et al. analyzed how Sim2Real transferability of AI models changed when trained on recursively generated data \cite{shumailov2024ai}. To further research in multi-modal sensing, Sim2Real transferability of multi-modal sensory signals was investigated, which demonstrated the significance of incorporating multi-modal sensing for Sim2Real stability \cite{chen2024general}. The aforementioned work in \cite{tobin2017domain}--\!\!\cite{chen2024general} mainly focuses on the impact of sensing on the Sim2Real transferability, demonstrating that large volumes of diverse simulation data that are aligned with real-world, and multi-modal data play a crucial role in reducing the real-world data usage. However, the research on the Sim2Real transfer in intelligent multi-modal sensing-communication integration is still lacking in the existing literature. As a result, conducting corresponding Sim2Real transfer to  validate the utility of digital-twin dataset is also urgently required.

However, constructing a new multi-modal sensing-communication digital-twin dataset with spatio-temporal consistency of both static objects and dynamic objects across real world and simulation environments, as well as validating the utility of the constructed digital-twin dataset, remain challenging. On one hand, restoring the dynamic objects to three-dimensional (3D) simulation environment from unlabeled real-world sensing data is challenging. That is, it is difficult to achieve spatial-temporal consistency of static objects and dynamic objects with the real world. On the other hand, there are challenges for cross-modal validation in Sim2Real transfer. At the data level, Sim2Real transfer in intelligent multi-modal sensing-communication integration involves heterogeneous modalities, including RGB images, depth maps, LiDAR point clouds, mmWave radar point clouds, and RF channel data. At the model level, multi-modal sensing–communication data differs significantly in acquisition frequency bands, data structures, and application orientation.

To fill these gaps, we consider a typical dynamic vehicle-to-infrastructure (V2I) scenario in 6G and construct a novel multi-modal sensing-communication digital-twin dataset, named SynthSoM-Twin. The SynthSoM-Twin dataset is an enhanced version of our previously proposed SynthSoM dataset~\cite{cheng2025synthsom}. Specifically, unlike the SynthSoM dataset~\cite{cheng2025synthsom}, the constructed SynthSoM-Twin dataset further contains multi-modal sensing–communication digital-twin data that is spatio-temporally consistent across the simulation environment and real world in DeepSense 6G dataset~\cite{alkhateeb2023deepsense}, i.e., a straight two-lane road on the south edge of Arizona State University and College Ave–5th St intersection in downtown Tempe. To validate the utility of SynthSoM-Twin dataset, we conduct Sim2Real transfer. The main contributions and novelties of this paper are outlined as follows.

\begin{figure*}
    \centering
    \includegraphics[width=1\linewidth]{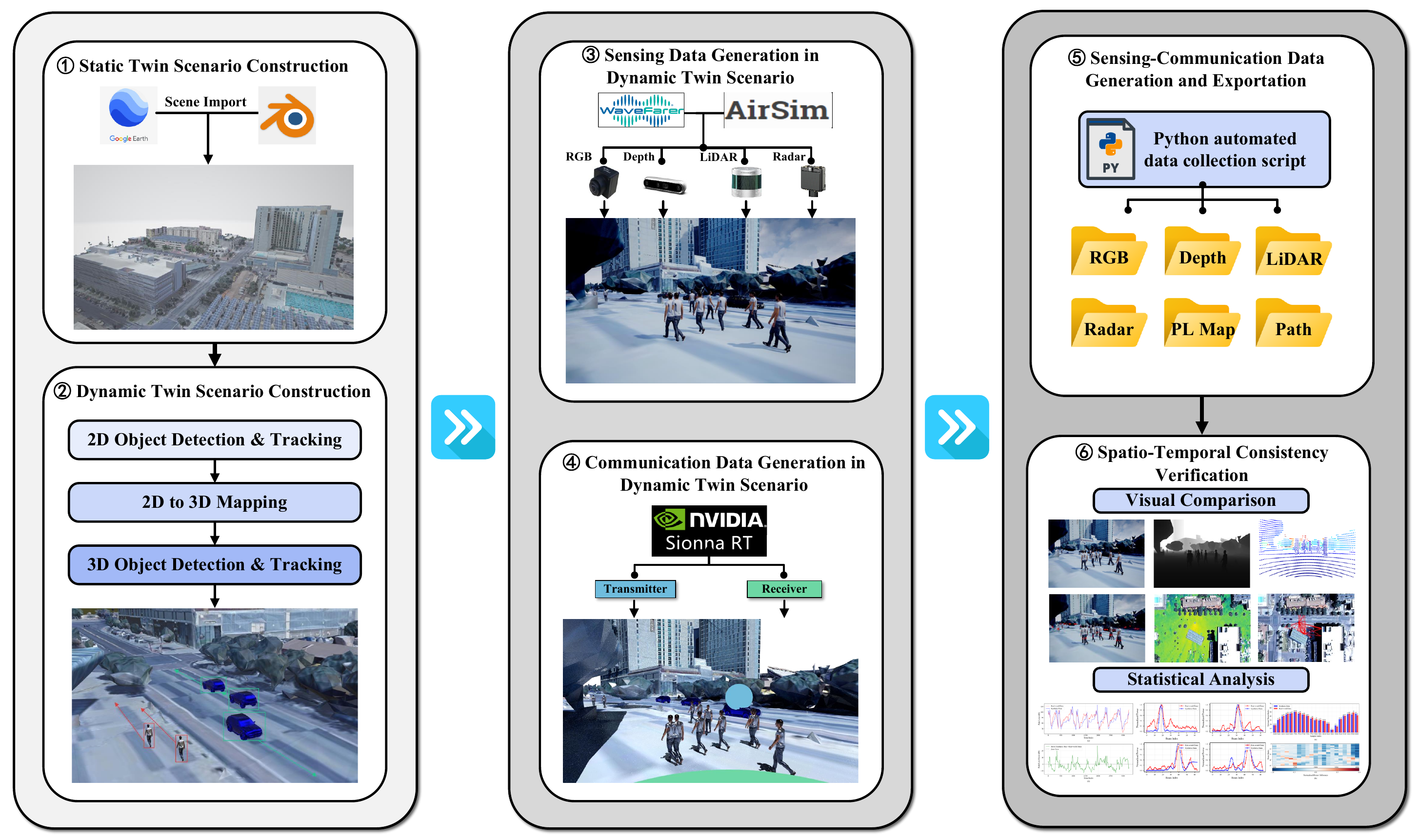}
    \caption{Framework of developing the proposed simulation platform for the construction of the SynthSoM-Twin dataset.}
    \label{fig: SynthSoM-Twin}
\end{figure*}
\begin{itemize}

\item In this paper, we construct a novel multi-modal sensing–communication dataset, named SynthSoM-Twin, which is spatio-temporally consistent with the real world. To construct the SynthSoM-Twin dataset, we propose a new framework that can extend the quantity and missing modality of existing real-world multi-modal sensing-communication dataset. 

\item For the static object, we exploit realistic 3D models captured in real world to restore static objects. For the dynamic object, as there is solely unlabeled real-world data in DeepSense 6G dataset~\cite{alkhateeb2023deepsense}, we exploit multi-modal sensing-assisted object detection and tracking algorithms that can restore the type and trajectory of dynamic objects. The constructed SynthSoM-Twin dataset contains spatio-temporally consistent data with spatio-temporal consistency of static objects and dynamic objects across simulation environments and real world, including 66,868 snapshots of synthetic RGB images, depth maps, LiDAR point clouds, mmWave radar point clouds, and large-scale and small-scale channel fading data.

\item To validate the utility of SynthSoM-Twin dataset through Sim2Real transfer, we select two representative cross-modal generation tasks, i.e., multi-modal sensing-assisted path loss generation and multi-modal sensing-assisted beam generation. Based on these two tasks, we design cross-modal generative models (CMGMs) that can generate path loss and beam power via multi-modal data to demonstrate the utility of SynthSoM-Twin dataset in reducing real-world data usage. 

\item Experimental results show that the SynthSoM-Twin dataset can reduce real-world data usage. The model training on the SynthSoM-Twin dataset achieves a proper practical performance. As the injection quantity of real-world data increases, the models' practical performance on two typical cross-modal generation tasks shows a trend of improvement. Injecting less than 15\% of real-world data can achieve a decent trade-off between real-world data usage and models' practical performance. This indicates that a proper digital-twin dataset can save over 85\% of the measurement cost on average.

\end{itemize}

The remainder of the paper is organized as follows. Section \ref{dataset} details the proposed framework and the construction of the SynthSoM-Twin dataset. Section \ref{CMGN} explains the design of two representative cross-modal generation tasks for Sim2Real transfer validation. Section \ref{result} shows the utility of the SynthSoM-Twin dataset and further explores the threshold of real-world data injection that can achieve a decent trade-off between real-world data usage and models' practical performance in multi-modal sensing-communication tasks. Finally, Section \ref{conclusion} summarizes the paper.

\section{SynthSoM-Twin: A Multi-Modal Sensing-Communication Digital-Twin Dataset}\label{dataset}

To construct a spatio-temporally consistent multi-modal sensing–communication dataset for reducing real-world data usage, we propose SynthSoM-Twin, i.e., a multi-modal sensing-communication digital-twin dataset. The objective of SynthSoM-Twin dataset is to provide a massive and high-quality data resource in which data generated from digital replica is similar to those observed in the real world, thus reducing real-world data usage. Towards this objective, it requires the geometric layout of static elements, e.g., roadways, building facades, and vegetation, and the kinematics of dynamic objects, e.g., vehicles and pedestrians, to be reproduced with high-level spatial accuracy and snapshot-level temporal synchrony. To construct the SynthSoM-Twin dataset, we propose a new framework that can extend the quantity and missing modality of existing real-world multi-modal sensing-communication dataset. The details of the framework are as follows. As shown in Fig.~\ref{fig: SynthSoM-Twin}, we first reconstruct a unified digital-twin scenario that merges static and dynamic content. The constructed digital-twin scenario is then imported into three high-fidelity simulators, including AirSim~\cite{shah2017airsim}, WaveFarer~\cite{remcom_wavefarer}, and Sionna RT~\cite{hoydis2023sionna}, to generate multi-modal information. AirSim generates non-RF sensing data, i.e., RGB images, depth maps, and LiDAR point clouds. WaveFarer generates RF sensing data, i.e., mmWave radar point clouds. Sionna RT generates RF channel data, i.e., path loss data, channel multipath data, and beam power data. A suite of Python automation scripts are used to orchestrate the aforementioned three simulators so that each simulation snapshot can be triggered by the original capture timestamps, ensuring that the $t^{\text{th}}$ snapshot in AirSim, WaveFarer, and Sionna RT describes the same physical instant. During execution, all sensor outputs are written to mode-specific folders that match the hierarchy of the real-world dataset. After the simulation, we conduct a two-stage spatio-temporal consistency validation. On one hand, visual overlays are leveraged to confirm pixel-level consistency between synthetic and real-world images or point clouds. On the other hand, statistical analysis similarity of empirical distributions of path loss, channel multipath information, and beam power to quantify residual divergence. The detailed construction steps of the SynthSoM-Twin dataset and validation results are given as follows.

\subsection{Static Twin Scenario Construction}

The static object of SynthSoM-Twin dataset is selected from the DeepSense 6G~\cite{alkhateeb2023deepsense} measurement campaigns, i.e., Scenario 31 and Scenario 32, as they represent two typical V2I scenarios, i.e., suburban and urban. Scenario 31 is a suburban scenario located at a straight two-lane road on the south edge of Arizona State University. Buildings are sparse, vegetation is low, and each snapshot contains few cars and pedestrians. Unlike Scenario 31, Scenario 32 is an urban scenario located at the College Ave–5th St intersection in downtown Tempe, where tall buildings and dense trees surround the crossroads. Furthermore, many vehicles start, slow, stop, and turn at the stop lines, and sidewalks are crowded with pedestrians, resulting in obvious blockage and multipath effects. Both scenarios contain precise GPS trajectories and RGB images for several intersections in Tempe, Arizona, USA. With the help of this information, we first delineate a rectangular geographic window that closely surrounds the whole scenario. The latitude–longitude coordinates of this window are then supplied to the blosm plugin in Blender, which requests the corresponding Google Earth 3D Tiles at a target level-of-detail of buildings with more details. This setting provides accurate mesh resolution, generating building models with facades resolved to approximately 10 cm and vegetation captured down to individual crown volumes. The imported tiles form a textured triangular mesh, whose vertex coordinates are initially represented as the center of the Earth, using a fixed Earth coordinate system with units in meters, and the same size as the physical scenario. To ensure a one-to-one correspondence between the digital and physical reference snapshots, we elevate the mmWave base station location to the origin of the local system, thus ensuring that $(x,y,z)=(0,0,0)$ in the twin is coincident with the physical transmitter (Tx) location.

Finally, the refined environment is exported simultaneously as an \texttt{.fbx} file, preserving materials for AirSim, as an \texttt{.stl} file, preserving materials for WaveFarer, and as an \texttt{.xml} description compliant with the Sionna RT, thus ensuring that both simulators ingest the same static model. The generating static object remains unchanged and provides the scenario foundation for the movement of dynamic objects in the next step.
\subsection{Dynamic Twin Scenario Construction}


In digital-twin scenarios, while high fidelity in static environments is crucial, the accurate reproduction of dynamic objects is equally indispensable. The underlying reason is that, moving targets, such as pedestrians and vehicles, not only affect the detection and tracking capabilities of sensors, but also alter the propagation characteristics of communication signals. Based on this, we focus on restoring the true motion trajectories of all traffic participants and precisely mapping them to the virtual world to ensure the consistency and high fidelity of the overall scenario in terms of spatio-temporal and electromagnetic characteristics.

Given that the DeepSense 6G dataset~\cite{alkhateeb2023deepsense} provides time-synchronized RGB images, LiDAR point clouds, and mmWave radar point clouds, we begin by classifying vehicles and pedestrians, which are two primary sources of electromagnetic scattering and occlusion in urban and suburban environments. Furthermore, we proceed step by step through three stages.

In the first stage, we extract the object type, two-dimensional (2D) position, and bounding box of dynamic objects from the real-world RGB images. The RGB images are first passed through a high-accuracy, pre-trained YOLOv12x~\cite{tian2025yolov12} detector, which outputs class-labeled 2D bounding boxes that contain dynamic objects, e.g., pedestrians and vehicles. To ensure temporal consistency, we then link the previous detections across snapshots using the BoT-SORT~\cite{aharon2022bot} tracker, which combines Mahalanobis-distance gating with deep appearance embeddings, which can achieve accurate tracking. The output of the tracker is a set of tracklets, each carrying a persistent identity, its per-snapshot image footprint, and an instantaneous pixel-velocity estimate derived from the tracker’s Kalman filter state vector.

In the second stage, we map RGB pixels to LiDAR points by aligning features from RGB images and LiDAR point clouds on the same plane, and supplement depth information for the previously generated 2D trajectory segments. First, we utilize the planar feature matching of the sensor-collected data as the initial attitude estimation to achieve a rough consistency of RGB images and LiDAR point clouds. Then, we refine the initial transformation by minimizing the photometric reprojection error of LiDAR point clouds on the image plane, further improving registration accuracy. Once the reprojection error is reduced to an acceptable range, we exploit a GPU-accelerated depth buffer (Z-buffer) algorithm to calculate the depth value of each pixel, establishing an accurate mapping between RGB pixels and LiDAR point clouds. Moreover, each pixel within the tracking box inherits the 3D coordinates $(x,y,z)$ of its corresponding nearest LiDAR point, thus extending the detection results from 2D to 3D.

In the third stage, we merge the previously extracted point cloud fragments into a complete set of raw 3D boundary points. Specifically, for each trajectory, we first filter out the subset of LiDAR point clouds that fall within the tracking contour after projection onto the image plane. We then utilize the torque clustering algorithm~\cite{yang2025autonomous}, a state-of-the-art algorithm that dynamically adjusts density thresholds based on actual point-to-point distances to ensure robust performance across both near-field and far-field ranges, to group this subset. As a consequence, we can separate noise point clouds from target object point clouds. For each retained cluster, we further construct a compact 3D bounding box that contains the actual dynamic object. The centroid, principal axis direction, and side length of this 3D bounding box respectively characterize the dynamic object's 3D pose in the current snapshot. Finally, to suppress occasional jitter caused by intermittent occlusion of the LiDAR sensor, we apply a sliding window smoother to the generated pose sequence, which can significantly improve the consistency and stability of the pose estimation.

Finally, the smoothed 3D trajectories of dynamic objects are exported in .txt format, containing the type of dynamic object, 3D coordinates, and direction, to facilitate subsequent import into AirSim, WaveFarer, and Sionna RT to drive the corresponding dynamic object movements. For vehicle and pedestrian models appearing in the real-world scenario, we select them from a catalog of vehicle and pedestrian models. Furthermore, the virtual vehicles and pedestrians in AirSim, WaveFarer, and Sionna RT are moving along the same trajectory, which ensures that the twin scenarios observed by the sensing twin and communication twin in each snapshot are not only visually similar to reality, but also consistent in terms of electromagnetic characteristics. This provides the necessary foundation for Sim2Real transfer validation in Section~\ref{result}.

\begin{figure*}
    \centering
    \includegraphics[width=1\linewidth]{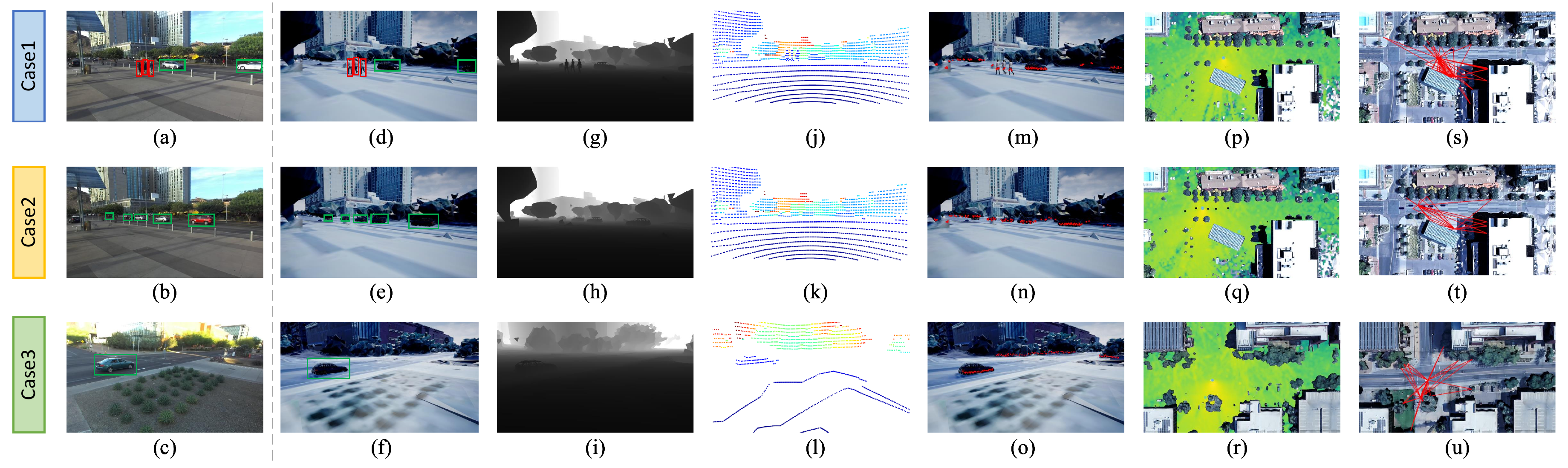}
    \caption{Qualitative cross-modal comparison between real-world data (left of the dashed line) and their synthetic counterparts in SynthSoM-Twin dataset (right).  Figs.~(a)--(c) show the real-world RGB images from three representative cases. Figs.~(d)--(f) show the synthetic RGB images from three representative cases. Figs.~(g)--(i) show the synthetic depth maps from three representative cases. Figs.~(j)--(l) show the synthetic LiDAR point clouds from three representative cases. Figs.~(m)--(o) show the synthetic mmWave radar point clouds from three representative cases. Figs.~(p)--(r) show the synthetic path loss map from three representative cases. Figs.~(s)--(u) show the synthetic ray tracing trajectories from three representative cases. Visual consistency across all modalities confirms snapshot-level geometric and photometric consistency between the synthetic data and real-world data.}
    \label{fig: SynthSoM-Twin_data_comparation}
\end{figure*}

\subsection{Multi-Modal Sensing Data Generation in Dynamic Twin Scenario}
Once the static mesh and a set of time-smoothed dynamic trajectories are finalized, both of them are imported into the AirSim simulator and WaveFarer simulator for sensing data ceneration. The \texttt{.fbx} file exported in Section~\ref{dataset}-A provides immovable geometry such as roads, building facades, vegetation, etc, while the dynamic object \texttt{.fbx} files and corresponding 3D trajectories generated in Section~\ref{dataset}-B provide the keyframe motion for each vehicle and pedestrian. The operation is also similar in WaveFarer. Since these two types of assets share the same coordinate system, no additional consistency is required during import. Simply playing a single global timeline can reproduce the real-world traffic flow with temporal consistency. 

Next, we need to deploy virtual sensors in the Unreal Engine to generate sensing data consistent with the measurement sensors. For RGB cameras and depth cameras, we leverage AirSim to create a virtual camera with parameters consistent with those of the actual camera. The pose of the camera is continuously adjusted in the simulation scenario to ensure that the preview image is consistent with the image captured by the actual camera. For the LiDAR sensor, we follow a similar camera setup process to configure it in AirSim. For the mmWave radar, we deploy the twin scenario in WaveFarer and configure it based on the measurement parameters and the pose of devices.

During the simulation process, the deployed virtual sensors generate sensing data at a given sampling rate.

\subsection{Communication Data Generation in Dynamic Twin Scenario}

The communication data generation in the dynamic twin scenario proceeds in parallel with the multi-modal sensing data generation whereas is guided by an entirely different fidelity metric, i.e., electromagnetic consistency, rather than photometric consistency. Next, we import the same static object model and dynamic object trajectories into Sionna RT, configure the radio materials and the Tx and receiver (Rx) antenna poses based on the installation height and viewing axis rotation angle provided in the measurement dataset, making fine adjustments to ensure the accuracy of the antenna poses~\cite{Hoydis2024ttt,cai2024switched}. In terms of antenna parameter settings, we follow the hardware parameters provided in the measurement dataset to maximize the reproduction of the antenna electromagnetic propagation characteristics.

During the simulation process, Sionna RT transmits a complete channel impulse response (CIR) tensor at a given sampling rate, and extracts path loss information, channel multipath information, and beam information.

\subsection{Sensing-Communication Data Generation and Exportation}
Leveraging the native logging utilities of AirSim, WaveFarer, and Sionna RT, we generate the sensing data and communication data in dynamic twin scenario with the same sampling rate, thus ensuring one-to-one temporal correspondence. For each timestamp, we generate the complete set of modalities that are relevant to the vehicle side and the roadside facilities. The RGB images are written as \texttt{.png} files, consisting of depth maps available in simulation utilize the same \texttt{.png} container, LiDAR point clouds are stored as \texttt{.npy} format, mmWave radar point clouds are exported as \texttt{.txt} format, channel multipath information is persisted in \texttt{.npy} format, and both beam power profiles and path loss traces are logged in \texttt{.txt}.

\subsection{Spatio-Temporal Consistency Verification}
Having produced the full synthetic dataset, we validate its fidelity by placing it side-by-side with the real measurements. Fig. \ref{fig: SynthSoM-Twin_data_comparation} shows three representative comparisons, where each row pairs the real-world data (left of the dashed line) with the synthetic data (right). Figs.~\ref{fig: SynthSoM-Twin_data_comparation}(a)--(c) show the real-world RGB images snapshot with 2D detection boxes overlaid. Figs.~\ref{fig: SynthSoM-Twin_data_comparation}(d)--(f) show the matched synthetic RGB images rendered by Unreal Engine. As can be seen from Figs.~\ref{fig: SynthSoM-Twin_data_comparation}(a)--(f), static objects, such as buildings, roads, and trees, remain nearly identical to the real-world scenario, while dynamic objects, such as pedestrians and vehicles, also maintain consistent positions and spatial occupancy with the real-world scenario. Figs.~\ref{fig: SynthSoM-Twin_data_comparation}(g)--(i) display the depth map available solely in the simulation, with its depth discontinuities consistent with the contours in Figs.~\ref{fig: SynthSoM-Twin_data_comparation}(a)--(c) and Figs.~\ref{fig: SynthSoM-Twin_data_comparation}(d)--(f). Figs.~\ref{fig: SynthSoM-Twin_data_comparation}(j)--(l) show the effect of converting the LiDAR point clouds of this snapshot to match the perspective of Figs.~\ref{fig: SynthSoM-Twin_data_comparation}(a)--(c). Comparing Figs.~\ref{fig: SynthSoM-Twin_data_comparation}(a)--(c) to Figs.~\ref{fig: SynthSoM-Twin_data_comparation}(j)--(l), it can be seen that the LiDAR point cloud also consists of consistent static objects and dynamic objects. Figs.~\ref{fig: SynthSoM-Twin_data_comparation}(m)--(o) overlays the mmWave radar detection results on the synthesized RGB image, indicating that the dynamic objects are placed in the correct distance range. The last two columns show the simulation results for the communication twin scenario, where Figs.~\ref{fig: SynthSoM-Twin_data_comparation}(p)--(r) and Figs.~\ref{fig: SynthSoM-Twin_data_comparation}(s)--(u) are the path loss map and channel multipath from a bird's-eye view. From Figs.~\ref{fig: SynthSoM-Twin_data_comparation}(p)--(u), it can be seen that the positions of the Tx and Rx are consistent with the real-world scenario. Additionally, Figs.~\ref{fig: SynthSoM-Twin_data_comparation}(p)--(u) depict that path loss and channel multipath are influenced by buildings, trees, and dynamic objects in the scenario. All these image comparison results demonstrate the spatio-temporal consistency of the dynamic twin scenario.

\begin{figure}
    \centering
    \includegraphics[width=1\linewidth]{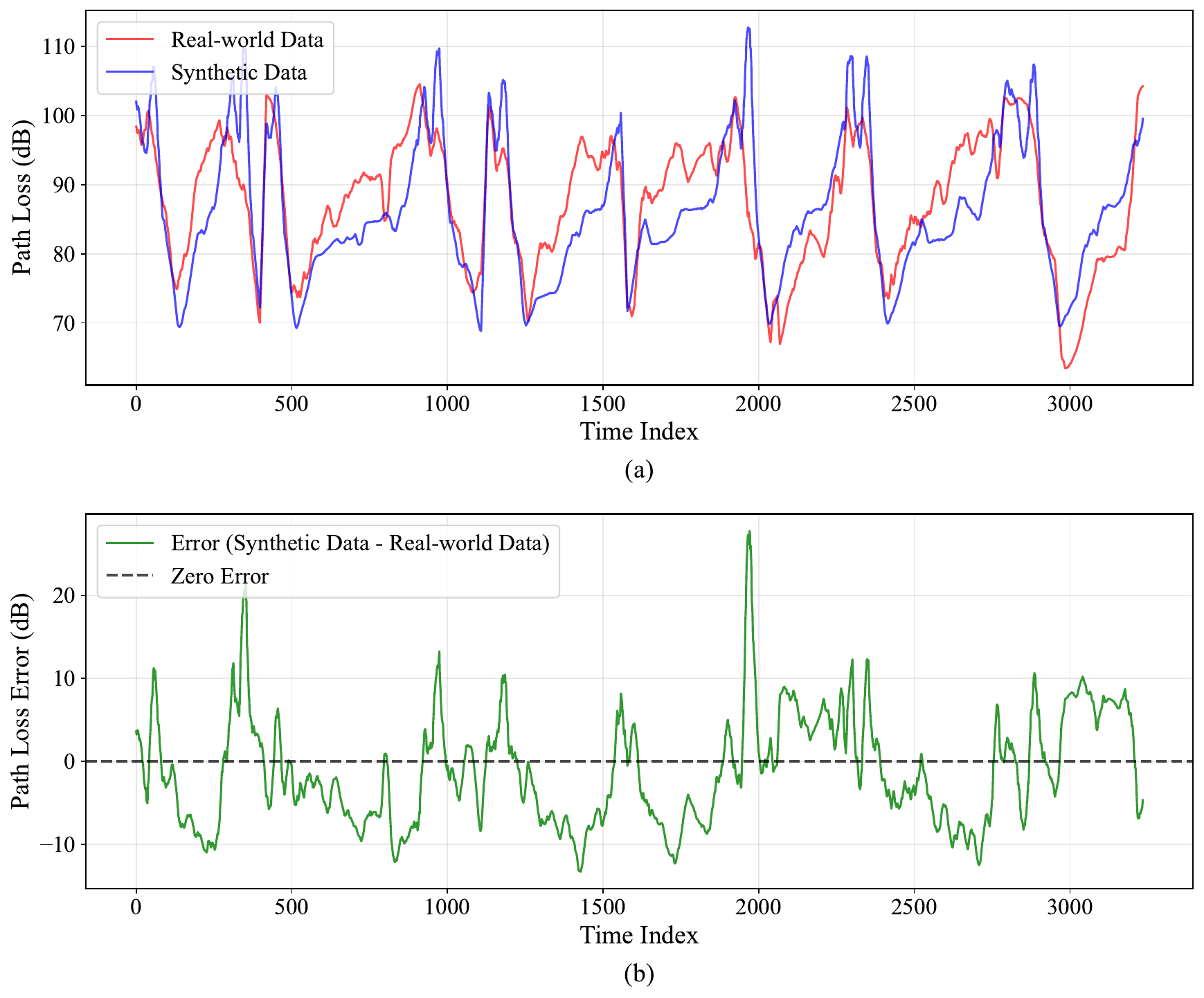}
    \caption{Temporal comparison of path loss. (a) Real-world path loss versus synthetic path loss. (b) Error between real-world path loss and synthetic path loss.}
    \label{fig:path_loss_comparison}
\end{figure}

\begin{figure}
    \centering
    \includegraphics[width=1\linewidth]{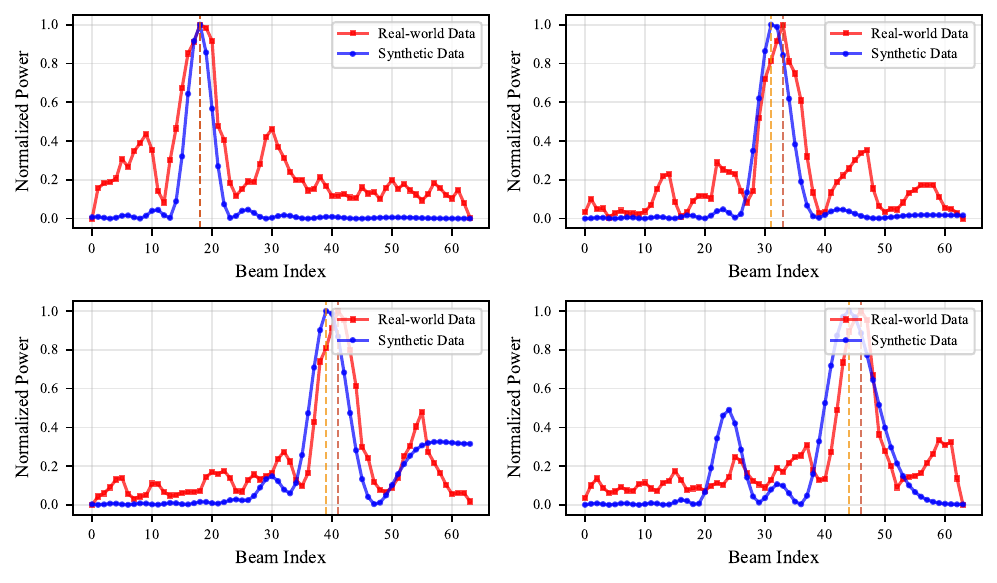}
    \caption{Frame-level beam power profiles for four timestamps, where the optimal-beam shift quantifies a tight correspondence between synthetic beam power and real-world beam power for each frame.}
    \label{fig:beam_power_local}
\end{figure}

\begin{figure}
    \centering
    \includegraphics[width=1\linewidth]{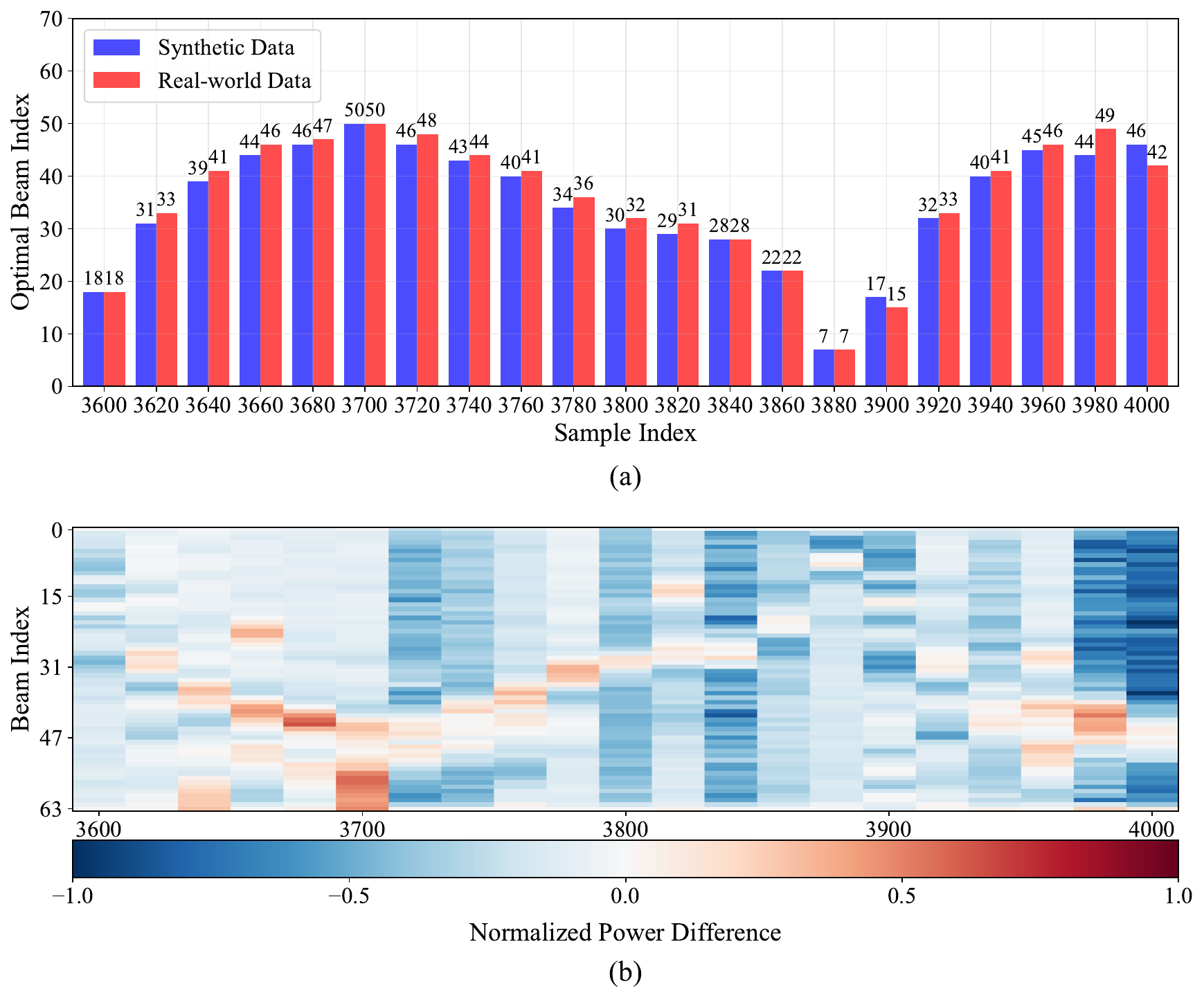}
    \caption{Global beam-forming statistics over 400 snapshots.  (a) Optimal beam-index histogram for simulation and measurement. (b) Heatmap of normalized power difference across all 64 beams, where light color denotes the close agreement and dark color highlights residual mismatch.}
    \label{fig:beam_power_global}
\end{figure}

The time series data in Fig. \ref{fig:path_loss_comparison} reveal that numerical values and time variation patterns of the synthetic path loss data are highly similar to the real-world path loss data. The fidelity of beam power data is corroborated in Fig.~\ref{fig:beam_power_local} and Fig.~\ref{fig:beam_power_global}. Specifically, Fig.~\ref{fig:beam_power_local} illutrates that snapshot-wise beam power spectra remain highly correlated and differ by no more than two codewords in optimal-beam index. The global statistics in Fig.~\ref{fig:beam_power_global} confirm that these consistency persist across 400 snapshots.  Therefore, Figs.~\ref{fig:path_loss_comparison}--\ref{fig:beam_power_global} verify that the constructed SynthSoM-Twin dataset achieves spatio-temporal consistency, which can support the Sim2Real transfer validation in Section~\ref{result}.

\begin{figure*}
    \centering
    \includegraphics[width=1\linewidth]{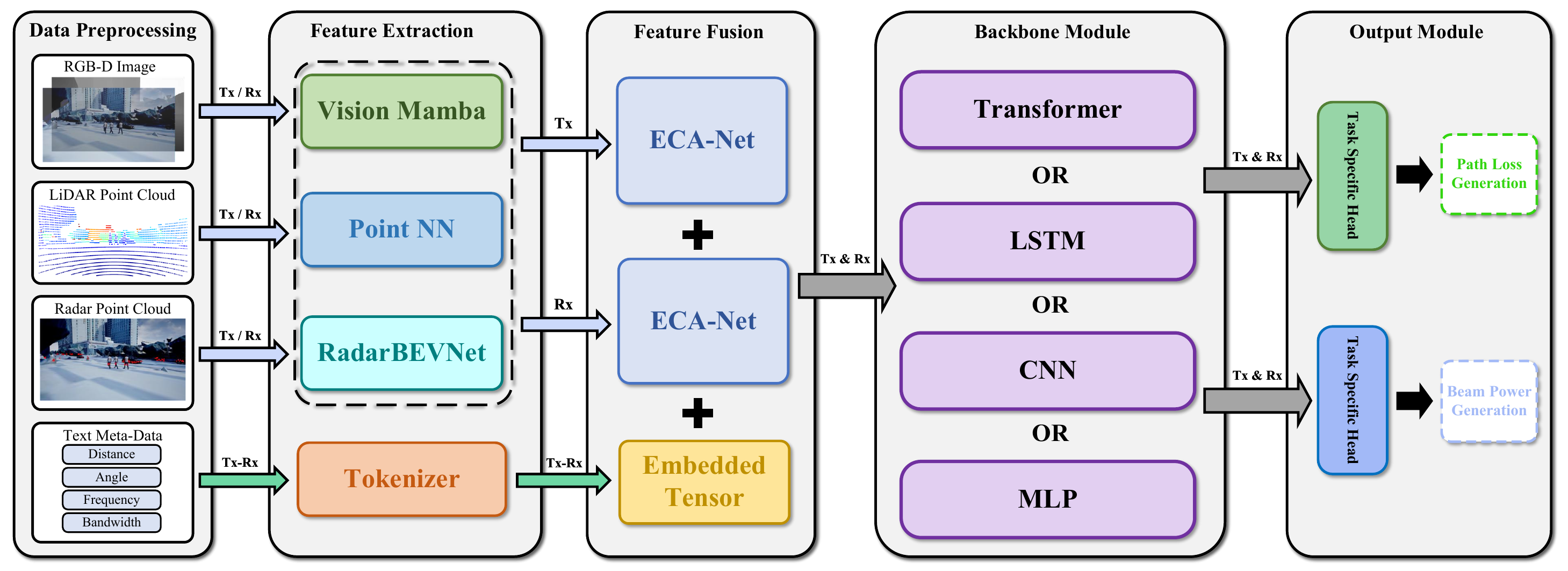}
    \caption{Architecture of the proposed CMGMs for multi-modal sensing-assisted path loss generation and multi-modal sensing-assisted beam generation.}
    \label{fig:pipeline}
\end{figure*}

\section{Cross-Modal Generation Task Design for Sim2Real Transfer Validation}\label{CMGN}

\subsection{Cross-Modal Generation Task Design}
To validate the utility of SynthSoM-Twin dataset, we conduct Sim2Real transfer by implementing two cross-modal downstream tasks using cross-modal generative models (CMGMs), i.e., cross-modal channel generation model and multi-modal sensing-assisted beam generation model. As shown in Fig.~\ref{fig:pipeline}, the composition of CMGM includes data processing, feature extractions, feature fusion, backbone module, and output module. The CMGM first preprocesses the two types of multi-modal input data, including sensing data collected from heterogeneous sensors and text meta-data that captures the link geometry. The generated sensing data includes RGB images, depth maps, LiDAR point clouds, and mmWave radar point clouds from Tx and Rx, while the text meta-data contains the distance and angles between Tx and Rx, carrier frequency of wireless channels, and communication bandwidth. 

After data preprocessing, we need to extract the features from the input data that reflect the dynamic states of the environment. Since the input data is heterogeneous, we need to efficiently extract features from the input data using different networks that adapt the structure of each input data. At the sensing side, three modality-specific encoders operate in parallel to extract the sensing data feature. RGB-D images that contain RGB images and depth maps are processed by a Mamba-Vision network~\cite{Zhu2024VisionMamba}, which is a token-mixing architecture that combines convolutional patch embedding with gated Mamba blocks, enabling efficient data extracting at modest compute cost. Point clouds sampled by the LiDAR sensors are routed through a Point NN backbone~\cite{zhang2023parameter} whose hierarchical set-abstraction blocks progressively aggregate local surface patches into a global representation that is permutation-invariant and robust to varying point density. Furthermore, mmWave radar point clouds are voxelized into a bird’s-eye tensor and passed to a RadarBEVNet~\cite{Lin2024RCBEVDet} that alternates sparse 3D convolutions with planar residual units to preserve range–azimuth structure and spatial sparsity. The feature maps emerging from these three branches are concatenated along the channel axis and compressed by an ECA-ResNet~\cite{Wang2020ECANet} fusion head, whose efficient channel-attention mechanism reweights modalities on-the-fly and projects the mixture to a fixed latent width, generating the sensing vector $\mathbf{p}$ that contains sensing features of Tx and Rx. In parallel, a tuple that contains the distance and angles between Tx and Rx, carrier frequency of wireless channels, and communication bandwidth is transformed by a two-layer multilayer perceptron with ReLU activations into the radio vector $\mathbf{m}$. 

After feature extraction and feature fusion, both feature vectors $\mathbf{p}$ and $\mathbf{m}$ are then linearly projected into a common hidden space. Each projection is prefixed with a learnable token identifier to preserve source identity, and the resulting pair of tokens is padded with a miniature sinusoidal position code. To eliminate the influence of the model in Sim2Real transfer, i.e., to focus on the constructed dataset itself, we design four widely used lightweight backbones for cross-modal generation tasks, including Transformer, LSTM, CNN, and MLP. The motivation for selecting Transformer, LSTM, CNN, and MLP is that they typify attention-based, recurrence-based, convolution-based, and MLP-based families, respectively, i.e., the classic architectural paradigms.

The fusion output, i.e., the hidden state of backbone model, is forwarded to a shared feed-forward trunk and diverges into two task-specific heads for multi-modal sensing-assisted path loss generation and multi-modal sensing-assisted beam generation. For multi-modal sensing-assisted path loss generation, the path loss value between Tx and Rx is treated as a scalar regression target~\cite{Wang2024ttt}. Therefore, we design the output head to generate path loss. A single sigmoid layer maps the hidden state to a scalar that is interpreted as path loss value in dB and trained under the loss function of mean-squared error (MSE), which is written as 
\begin{equation}
\mathcal{L}_{\text{PL}}=\frac1N\sum_{i=1}^{N}\!\bigl(\hat{L}_{i}-L_{i}\bigr)^{2}
\end{equation}
where \(N\) is the batch size, \(i\) indexes samples, \(\hat{L}_{i}\) is the generated path loss, and \(L_{i}\) is the ground truth.

For multi-modal sensing-assisted beam generation, it models directional beam selection as a $64$-dimensional vector regression problem, where the model needs to reproduce the normalized power $\hat{\mathbf{b}}\in\mathbb{R}^{64}$ on a predefined codebook. To minimize the error of the generated beam power while maximizing the classification of the maximum beam power index, we use Top-$k$ cross-entropy as the loss function of beam power generation, which is given by  

\begin{equation}
\mathcal{L}_{\text{BP}}=-\frac1N\sum_{i=1}^{N}\!
\mathbf{M}_{k}\!\bigl(\mathbf{b}_{i}\bigr)\cdot
\log\!\bigl[\text{softmax}(\hat{\mathbf{b}}_{i}/\tau)\bigr]
\end{equation}
where \(N\) is the batch size, \(i\) indexes samples, and \(\mathbf{b}_{i}\in\mathbb{R}^{64}\) denotes the ground truth. A \(\text{softmax}(\cdot)\) is applied across the beam dimension and \(\mathbf{M}_{k}(\mathbf{b}_{i})\in[0,1]^{64}\) is a Sinkhorn-derived soft Top-\(k\) label that concentrates probability mass on the \(k\) strongest beams and is normalized to sum to one. Furthermore, \(k\) is an integer hyperparameter and \(\tau>0\) controls the sharpness of the output distribution.

Following the steps from data preprocessing to generating output, the CMGM achieves the cross-modal generation of communication data from easily accessible sensing data by exploring the potential mapping mechanism between sensing and communications. More importantly, the developed CMGM can be exploited to validate the utility of SynthSoM-Twin dataset in reducing real-world data usage.

\subsection{Sim2Real Transfer Validation Strategy}
To validate the utility of SynthSoM-Twin dataset through Sim2Real transfer via the CMGM, we exploit three approaches, i.e., train on synthetic and test on real (TSTR), train on real and test on real (TRTR), and fine-tune (FT). In the TSTR approach, the CMGM is trained solely on the SynthSoM-Twin training set, which contains synthetic RGB images, depth maps, LiDAR point clouds, mmWave radar point clouds, path loss, as well as beam power. The CMGM is optimized to generate either a path loss value or a 64-dimensional beam power based on the sensing input. Then, the pre-trained CMGM is tested on the divided test set in the real-world DeepSense 6G dataset~\cite{alkhateeb2023deepsense} to demonstrate the basic performance of model solely utilized SynthSoM-Twin dataset. In the TRTR approach, the CMGM is trained solely on the DeepSense 6G ~\cite{alkhateeb2023deepsense} training set, which contains real-world RGB images, LiDAR point clouds, mmWave radar point clouds, path loss, and beam power. Then, the pre-trained CMGM is tested on the same test set to demonstrate the basic performance of model solely utilized DeepSense 6G dataset~\cite{alkhateeb2023deepsense}. To further explore the threshold of real-world data injection that can achieve a decent trade-off between real-world data usage and models' practical performance, in the FT approach, we consider pre-train a model on massive synthetic data and then fine-tune it with a small amount of real-world data. Specifically, we fine-tune the CMGM pre-trained on the completed synthetic dataset by injecting $s\in \{10,20,\dots,400,500,600\}$) continuous real-world data segments as mini batches. The reason we maintain the temporal continuity of the fragments is that real-world data collection is typically collected without gaps, and we aim to preserve their temporal statistical properties. In the FT approach, the task-specific heads are tuned to adapt to the real-world domain.

\section{Experimental Results}\label{result}

\subsection{Experimental Setup and Evaluation Metrics}
\label{subsec:setup}
In the experiments, we consider four backbone models, including Transformer, LSTM, CNN, and MLP, while keeping other components of the CMGM unchanged. Furthermore, we consider three cross-modal generation approaches, including TSTR, TRTR, and FT. To quantify the performance of CMGM, several metrics are utilized for different tasks. For path loss generation, we measure root-mean-square error (RMSE) and mean absolute error (MAE) to quantify the quality of the generated path loss value. RMSE emphasizes large errors and tail risk, while MAE reflects typical absolute deviation and robustness to outliers. For beam power generation, we select Top-1 accuracy and Top-3 accuracy of the generated max power beam index. The reason for including Top-3 accuracy is that Top-3 accuracy reflects the fact that adjacent beams often offer comparable gain. We utilize AdamW optimizer with a fixed learning rate of $1\times10^{-3}$, and further the batch size to 4 and training epochs to 12. The detailed hyper-parameters are summarized in Table~\ref{tab:hyper}.

\begin{table}[t]
    \centering
    \caption{Key hyper-parameters in experiments.} 
    \label{tab:hyper}
    \begin{tabular}{|c|c|}
        \hline
        \textbf{Hyperparameter}      & \textbf{Setting}                           \\ \hline
        Batch size                   & 4                                         \\ \hline
        Epochs                       & 12                                        \\ \hline
        Optimizer                    & AdamW                                      \\ \hline
        Learning rate       & $1.0\times10^{-3}$                           \\ \hline
    \end{tabular}
\end{table}

To validate the utility of SynthSoM-Twin dataset and further explore the threshold of real-world data injection that can achieve a decent trade-off between real-world data usage and models' practical performance, we setup four scenarios with different vehicular traffic densities (VTDs), including \textit{urban scenario with consistent VTD}, \textit{suburban scenario with consistent VTD}, \textit{urban scenario with inconsistent lower VTD}, and \textit{urban scenario with inconsistent higher VTD}. Specifically, the \textit{urban scenario with consistent VTD} refers to an urban scenario at a 60 GHz carrier frequency with 2 GHz communication bandwidth. The scenario is based on DeepSense 6G \textit{Scenario 32}~\cite{alkhateeb2023deepsense}, which is located at the College Ave–5th St intersection in downtown Tempe with tall buildings and dense trees surrounding the crossroads. As the dynamic objects are restored from DeepSense 6G \textit{Scenario 32}~\cite{alkhateeb2023deepsense}, the simulated VTD is equal to the real-world VTD with the consistent number and trajectory of vehicles. Based on the scenario, we conduct Sim2Real transfer on multi-modal sensing-assisted path loss generation. The \textit{suburban scenario with consistent VTD} refers to a suburban scenario at a 60 GHz carrier frequency with 2 GHz communication bandwidth. The scenario is based on DeepSense 6G \textit{Scenario 31}~\cite{alkhateeb2023deepsense}, which is located on a straight two-lane road on the south edge of Arizona State University with sparse buildings, low vegetation, few cars and almost no pedestrians. As the dynamic objects are restored from DeepSense 6G \textit{Scenario 31}~\cite{alkhateeb2023deepsense}, the simulated VTD is equal to the real-world VTD with the consistent number and trajectory of vehicles. Based on the scenario, we conduct Sim2Real transfer for multi-modal sensing-assisted beam generation, i.e., beam power generation. To validate the importance of achieving spatio-temporal consistency of static objects and dynamic objects across real world and simulation environment, we further setup two scenarios that have consistent static objects of DeepSense 6G \textit{Scenario 32}~\cite{alkhateeb2023deepsense} and inconsistent simulated VTD. Specifically, \textit{urban scenario with inconsistent lower VTD} has lower simulated VTD compared to real-world VTD, and \textit{urban scenario with inconsistent higher VTD} has higher simulated VTD compared to real-world VTD. 


\begin{figure}[t]
    \centering
    \includegraphics[width=0.8\linewidth]{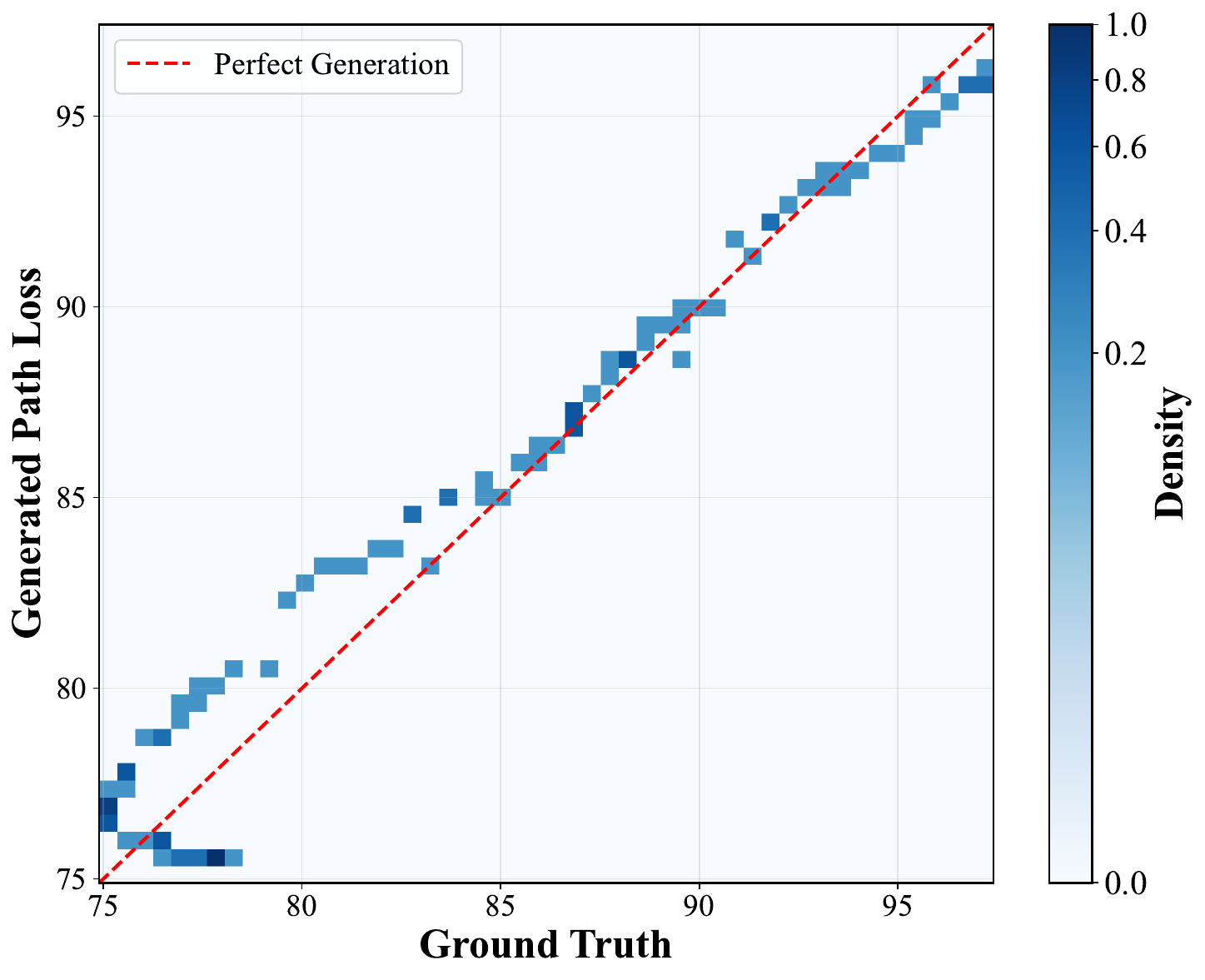}
    \caption{Distribution of generated path loss versus ground truth.}
    \label{fig:pl_heat}
\end{figure}


\begin{figure}[t]
    \centering
    \includegraphics[width=0.8\linewidth]{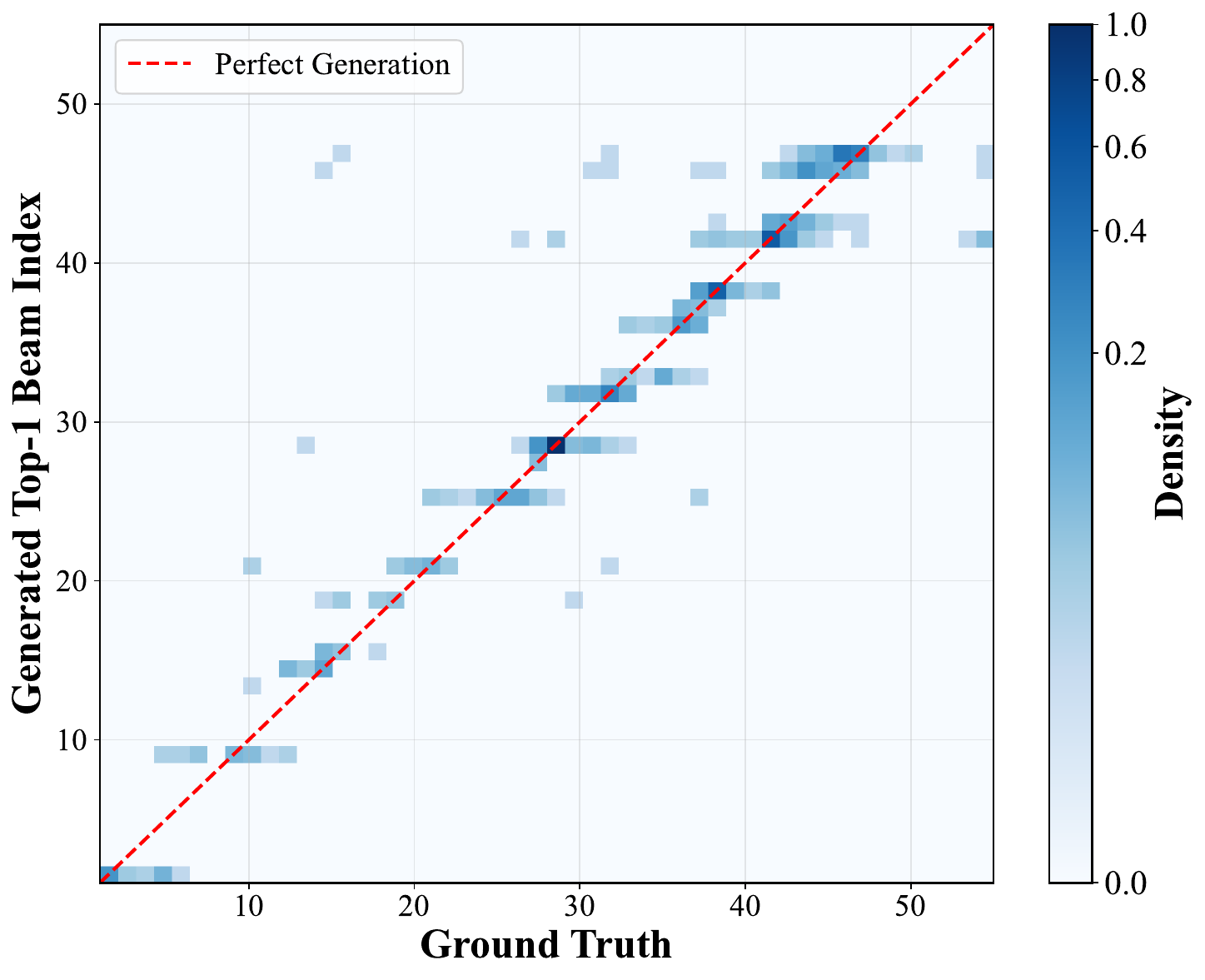}
    \caption{Distribution of generated Top-1 beam index versus ground truth.}
    \label{fig:bp_heat}
\end{figure}


\subsection{Effectiveness Evaluation on Cross-Modal Generative Model}
\label{subsec:pl}
To validate the effectiveness of CMGM in supporting cross-modal generation tasks for Sim2Real transfer, we evaluate the practical performance of CMGM on path loss generation task and beam power generation task via TRTR approach.
We first evaluate the practical performance of CMGM in path loss generation task under \textit{urban scenario with consistent VTD} in SynthSoM-Twin dataset. Fig.~\ref{fig:pl_heat} visualizes the joint distribution of generated path loss and ground truth.
As shown in Fig.~\ref{fig:pl_heat}, the 2D density heatmap of generated path loss versus ground truth concentrates tightly along $y=x$ line across the 75–100 dB range, indicating close agreement between the distribution of generated path loss and ground truth. Overall, the results show that the CMGM have decent practical performance on path loss generation task, which can support fundamental path loss generation tasks.
We then evaluate the practical performance of CMGM in beam power generation task under \textit{suburban scenario with consistent VTD} in SynthSoM-Twin dataset. Fig.~\ref{fig:bp_heat} visualizes the joint distribution of the generated Top-1 beam index and ground truth. The density concentrates along the $y=x$ diagonal with a narrow band, indicating close agreement between the distribution of generated Top-1 beam index and ground truth. Overall, the results show that the CMGM have decent practical performance on beam power generation task, which can support multi-modal sensing-assisted downstream tasks. As a result, the designed CMGM is effective, and training and validation using real-world data directly has a certain level of accuracy.



\subsection{Sim2Real Transfer Validation}
\label{subsec:val}

\begin{figure}
    \centering
    \includegraphics[width=1\linewidth]{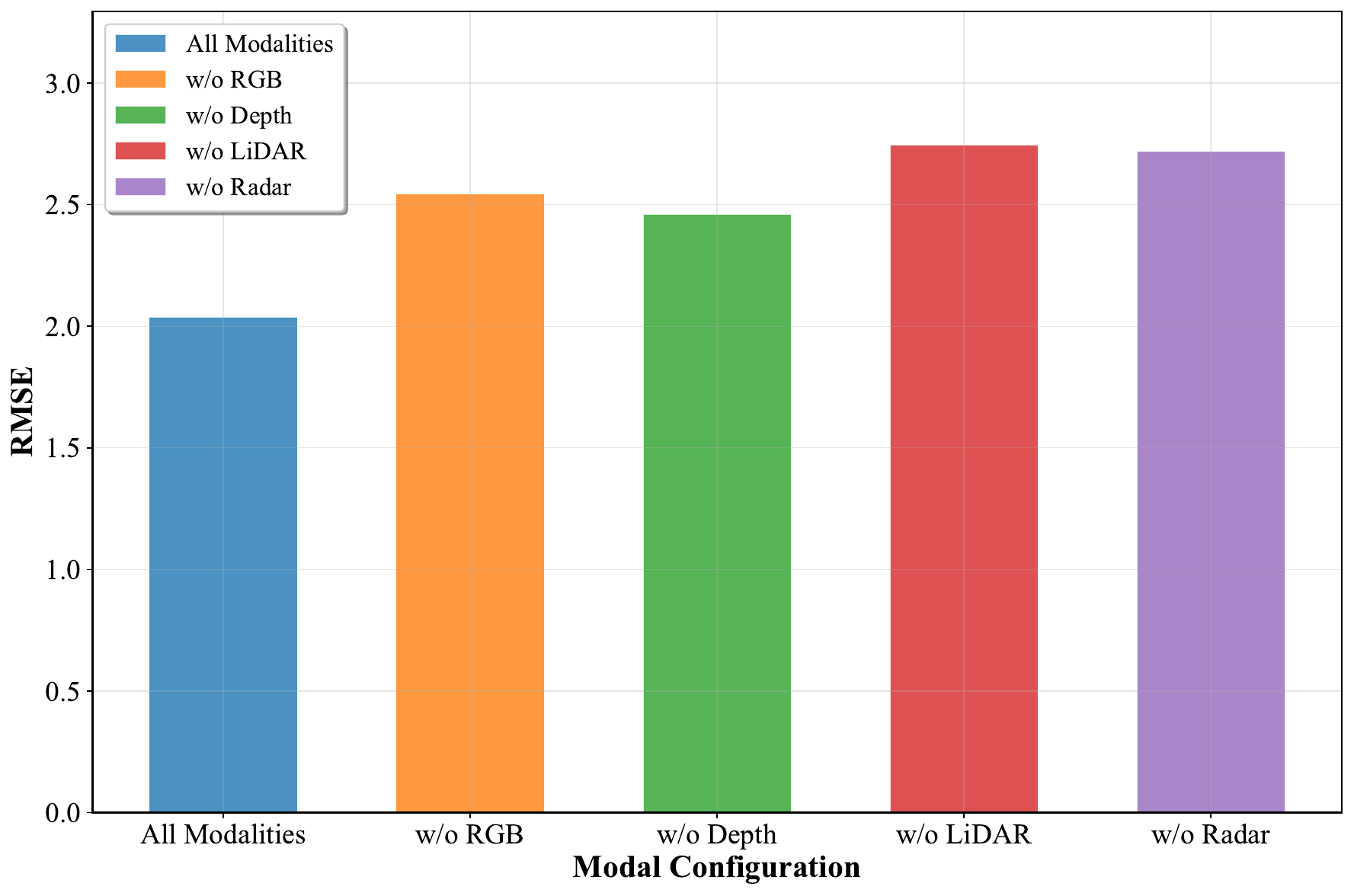}
    \caption{Models' practical performance under different numbers of Modality input.}
    \label{fig:modal}
\end{figure}

\noindent\textit{1) Impact of multi-modal inputs on models' practical performance:} To validate the importance of using multi-modal data in SynthSoM-Twin dataset for cross-modal generation tasks, we setup several baselines by excluding one of the modal inputs from SynthSoM-Twin dataset, including w/o RGB, w/o Depth, w/o LiDAR, and w/o Radar. As shown in Fig.~\ref{fig:modal}, missing of each modality input reduces models' practical performance, demonstrating the necessity of multi-modal inputs. Note that, inputting depth map modalities not present in the DeepSense 6G dataset~\cite{alkhateeb2023deepsense} will enhance model's practical performance. This indicates that even though depth map is unavailable in real-world measurements, the additional synthetic depth map in SynthSoM-Twin can make the model better capture the underlying geometric structure and cross-modal correlations, enabling the model to learn representations that are harder to obtain from the measured modalities alone. Therefore, it is necessary to include extensive multi-modal data, including synthetic-only modalities such as depth, in the SynthSoM-Twin dataset.

\begin{figure}
    \centering
    \includegraphics[width=1\linewidth]{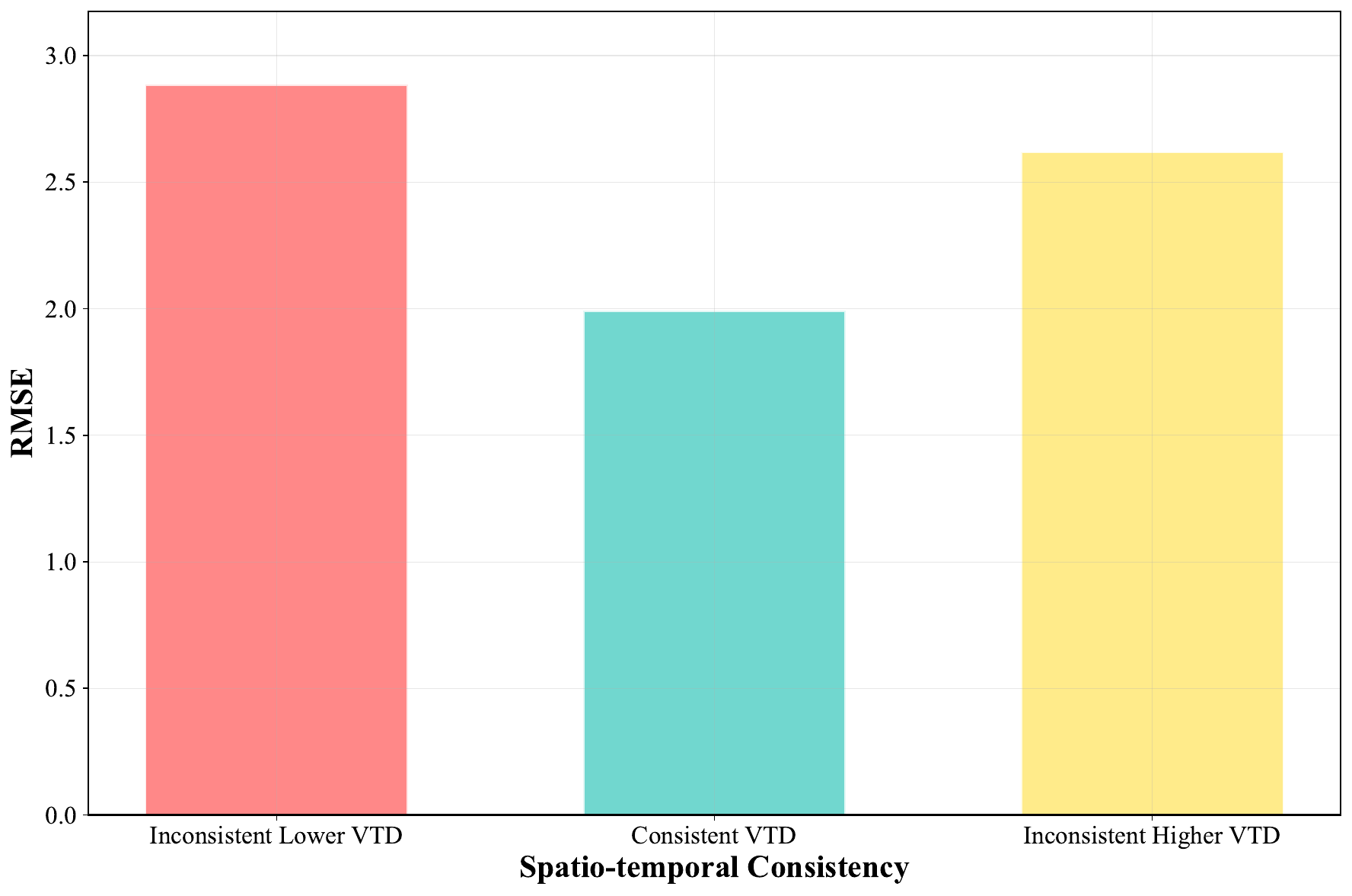}
    \caption{Models' practical performance under varying degrees of spatio-temporal consistency in dynamic objects.}
    \label{fig:st-consist}
\end{figure}

\noindent\textit{2) Effect of spatio-temporal consistency between simulation environment and real world:} To validate the importance of achieving spatio-temporal consistency of static objects and dynamic objects across real world and simulation environment, we compare the performance of model training on \textit{urban scenario with consistent VTD} with \textit{urban scenario with inconsistent lower VTD} and \textit{urban scenario with inconsistent higher VTD}. Specifically, Consistent VTD refers to the urban scenario with consistent VTD that achieves spatio-temporal consistency in both static objects and dynamic objects across real world and simulation environments. Inconsistent Lower VTD refers to the urban scenario with inconsistent lower VTD. Inconsistent Higher VTD refers to the urban scenario with inconsistent higher VTD. As shown in Fig.~\ref{fig:st-consist}, whether Inconsistent Lower VTD or Inconsistent Higher VTD, they both have a lower  practical performance than Consistent VTD, showing that using synthetic datasets without achieving spatio-temporal consistency of static and dynamic objects reduce models' practical performance. Therefore, it is important to achieve spatio-temporal consistency of static and dynamic objects across real world and simulation environment when constructing digital-twin dataset.


\begin{figure}
    \centering
    \includegraphics[width=1\linewidth]{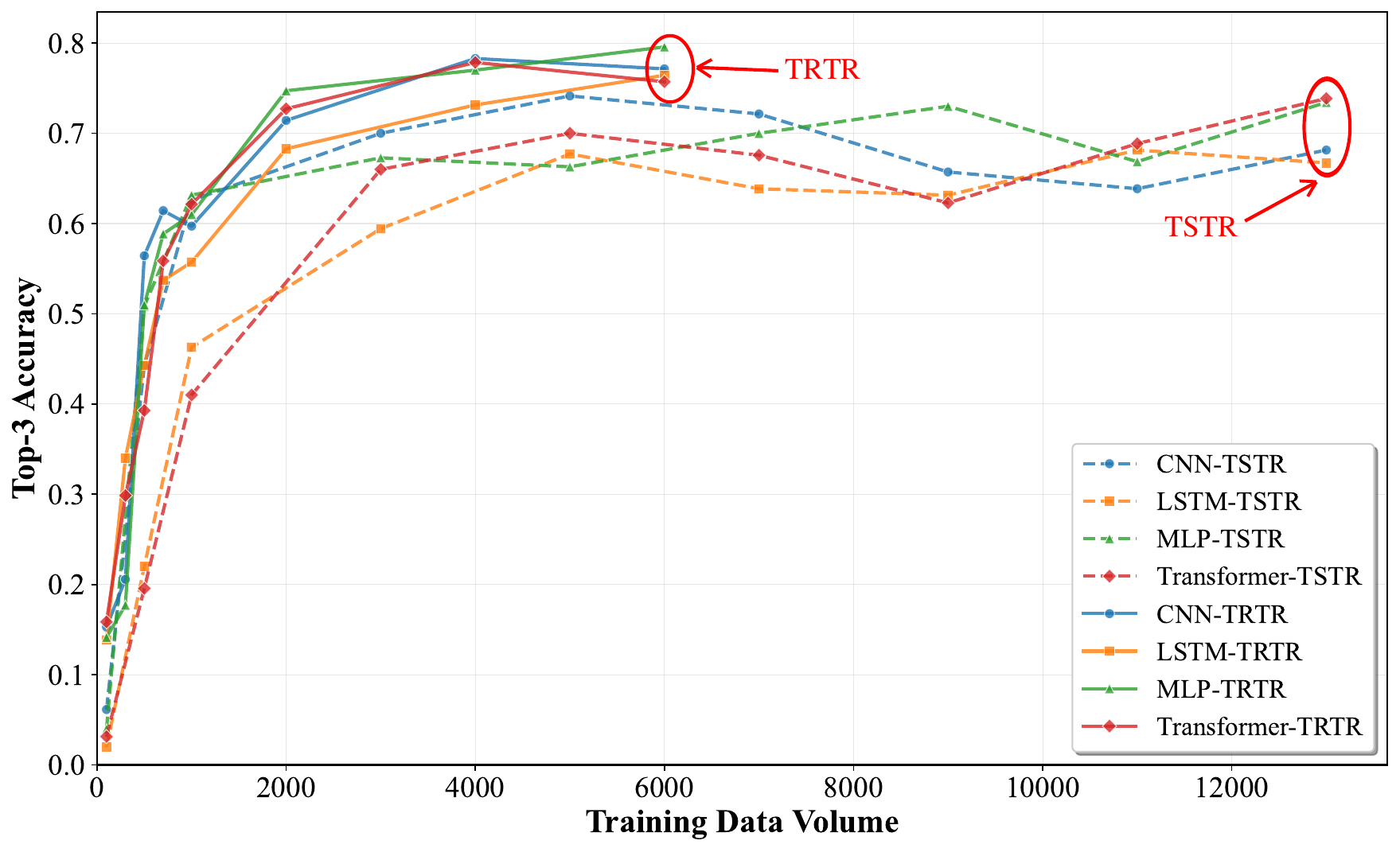}
    \caption{Models' practical performance under different training quantity of synthetic/real-world data.}
    \label{fig:quantity}
\end{figure}

\noindent\textit{3) Utility of SynthSoM-Twin under varying training data quantities:} To validate the utility of the SynthSoM-Twin dataset, we evaluate the models' practical performance under different training quantity of synthetic/real-world data in \textit{suburban scenario with consistent VTD}. The TRTR was trained using 6,000 snapshots of real-world data, while TSTR was trained on a synthetic dataset augmented to 13,700 snapshots to fully explore the advantages of synthetic data volume. As shown in Fig.~\ref{fig:quantity}, for both TSTR and TRTR, as the training quantity of synthetic/real-world data increases, the models' practical performance shows a trend of improvement, which meets the scaling law. We can also observe that although TSTR has lower practical performance than TRTR, both TSTR and TRTR exhibit similar practical performance and maintain high practical performance, thus validating the utility of the SynthSoM-Twin dataset. 

\begin{figure}
    \centering
    \includegraphics[width=1\linewidth]{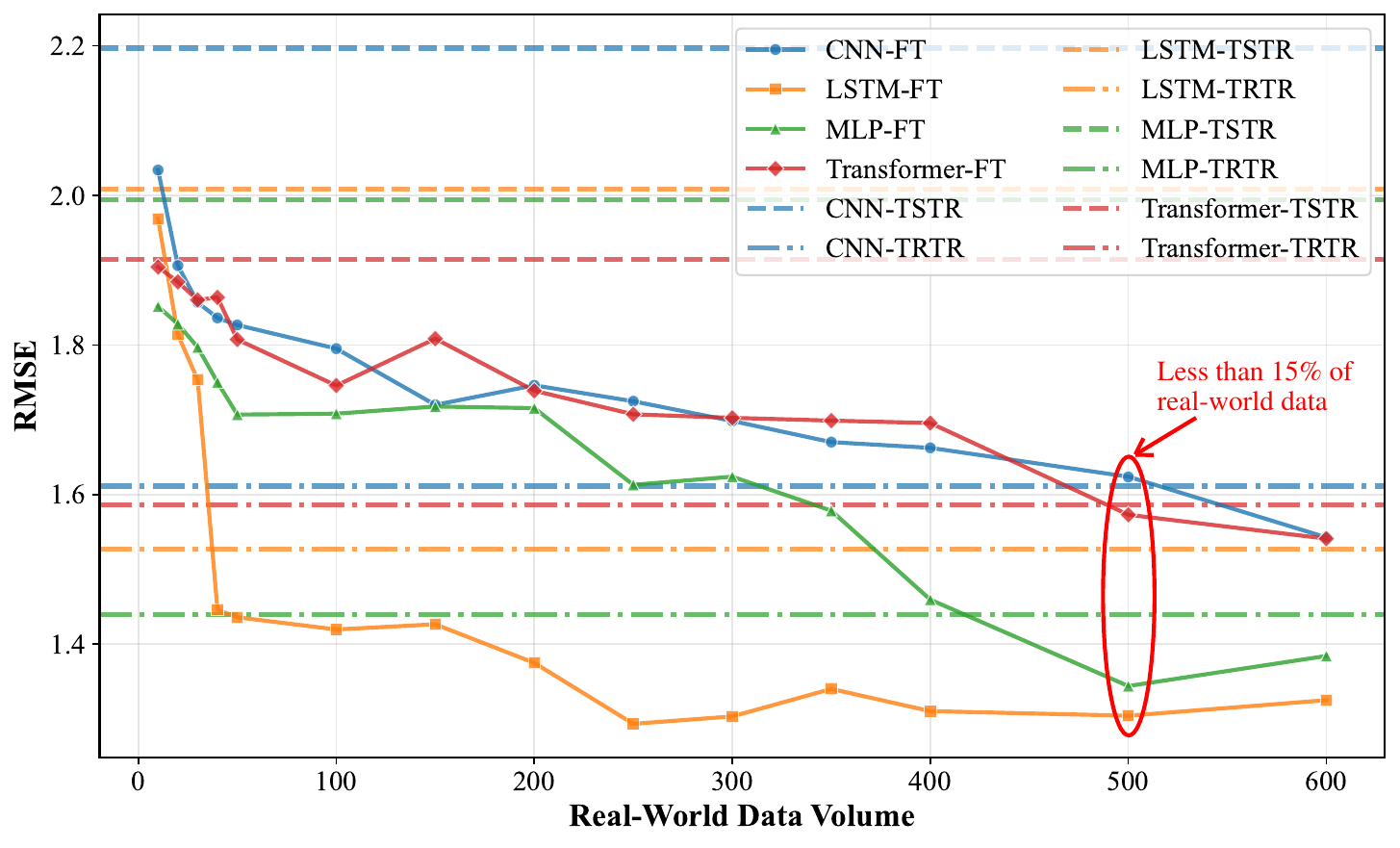}
    \caption{The practical performance of model fine-tuned with different quantities of real-world data.}
    \label{fig:inj_quantity}
\end{figure}

\noindent\textit{4) Balancing models' practical performance and real-world data usage:} Furthermore, we explore the threshold of real-world data injection that can achieve a decent trade-off between real-world data usage and models' practical performance through FT approach. In the FT approach, we fine-tune the CMGM pre-trained on \textit{suburban scenario with consistent VTD} by injecting continuous real-world data segments that covering $s\in \{10,20,\dots,400,500,600\}$. As shown in Fig.~\ref{fig:inj_quantity}, a small amount of real-world data injection can significantly improve models' practical performance, i.e., the performance deployed in the real world. Additionally, around 500 volume of real-world data, i.e., less than 15\% of real-world data, can make the four backbone models reach a similar practical performance compared to TRTR. This indicates that a proper digital-twin dataset can save over 85\% of the measurement cost on average. 

\section{Conclusion}\label{conclusion}
In this paper, we have constructed SynthSoM-Twin, a multi-modal sensing-communication digital-twin dataset for Sim2Real transfer via SoM. To construct the SynthSoM-Twin dataset, we have proposed a new framework that can extend the quantity and missing modality of existing real-world multi-modal sensing-communication dataset. Through multi-modal sensing-assisted object detection and tracking algorithms, the SynthSoM-Twin dataset has achieved spatio-temporal consistency of static objects and dynamic objects across real world and simulation environments. To validate the utility of the SynthSoM-Twin dataset, we have conducted Sim2Real transfer by implementing two cross-modal downstream tasks via CMGMs, i.e., cross-modal channel generation model and multi-modal sensing-assisted beam generation model. Furthermore, we have explored the threshold of real-world data injection that can achieve a decent trade-off between real-world data usage and models' practical performance. Experimental results have shown that the model training on the SynthSoM-Twin dataset has achieved a proper practical performance, and the injection of real-world data has further facilitated Sim2Real transferability. Based on the SynthSoM-Twin dataset, injecting less than 15\% of real-world data has achieved a decent trade-off between real-world data usage and models' practical performance. This has indicated that a proper digital-twin dataset can save over 85\% of the measurement cost on average.

\bibliographystyle{IEEEtran}

\bibliography{main}

\end{document}